\magnification=1200
\vsize=8.5truein
\hsize=6truein
\baselineskip=20pt

\def\R{{\bf R}} 
\def\Z{{\bf Z}} 
\def\wt{\widetilde} 
\def\eps{\varepsilon}
\def\prt{\partial}

\centerline{\bf Configurational transition in a 
Fleming-Viot-type model  }
\centerline{\bf and probabilistic
interpretation of Laplacian eigenfunctions} 
\centerline{by}
\centerline{Krzysztof Burdzy$^1$, Robert Ho{\l}yst$^2$, 
David Ingerman$^1$ and Peter March$^3$}
\vskip 20pt
\centerline{ $^1$Department of Mathematics, University of Washington, 
Seattle, WA 98195,USA}
\centerline { $^2$Institute of Physical Chemistry of the
Polish Academy of Sciences and College of Sciences,} 
\centerline{Dept. III, 
Kasprzaka 44/52, 
01224 Warsaw, Poland}
\centerline{ $^3$ Department of Mathematics, Ohio State University,
Columbus, OH 43210, USA}
\vskip 20pt
\centerline{\bf Abstract}
We analyze and simulate a two dimensional Brownian multi-type 
particle system
with death and branching (birth) depending on the 
position of 
particles of different types. 
The system is confined in the two dimensional box,
whose boundaries act as the sink of Brownian particles. 
The branching rate matches the
death rate so that the total
number of particles is kept constant.
In the case of $m$ types of
particles in the rectangular box of size $a,b$ and elongated shape
$a\gg b$ we observe that the stationary distribution of particles corresponds
to the $m$-th Laplacian eigenfunction. 
For smaller elongations $a>b$ we find a 
configurational transition to a new limiting distribution. The ratio $a/b$
for which the transition occurs is related to the value of the 
$m$-th eigenvalue of the Laplacian with rectangular boundaries.
 
\vfill\eject
\centerline{\bf I Introduction}

It is remarkable how the simple systems with few deterministic
rules (such as Life [1] or cellular automata [2])
can generate complex structures. In the population dynamics
however one often uses stochastic  models. For example 
as has been recently shown [3] the addition of 
stochastic factor into the Life game
favors diversity of structures, contrary to the
original model in which diversity is the decreasing function of time.
Introduction of the probabilistic factor in the cellular automata
in the description of the dynamics of the social impact in the population
[4] leads to the
complex spatial and temporal intermittent behavior.  
In the genome population dynamics [5-6] 
one uses stochastic processes
such as super-Brownian motion
or Fleming-Viot processes. The model presented in this
paper is a special type of population dynamic stochastic model.

The dynamics of systems with two competing species has been studied
with the emphasis on the spatial heterogeneity influence on the
temporal evolution and spatial organization [7-9]. In the case of strong
competition only one of the two survives, which means that 
the average lifetime of two species can be different. 
In our model we are concerned
with the spatial distribution
of three or more competing  species. Our model differs from those mentioned 
above because we study the case when
the total number of particles is constant and the 
average lifetime of different species is the same in the long time limit.

Here we describe
a behavior of a population of $m$ different types of particles with
Brownian dynamics confined in the two dimensional box, whose walls act
as sinks for the particles. Additionally we assume that if two particles
of different type occupy the same lattice point, both are killed.
The birth rules are chosen in such a way as to keep the 
number of particles constant at each time step and to ensure that
the average lifetime of any type of particles is the same in the
long time limit.
As an example we show (Fig.1)
the stationary
configurations for 3 types of particles in a rectangular box
of size $a \times b$.
For $a/b>1.63$ the particles of different types occupy domains
of rectangular shapes (Fig.1a). 
We call such configuration elementary and show (in section V)
that it is related to the third Laplacian eigenfunction. When the size ratio
decreases below 1.63 the configuration changes its character as shown in
Figs.1bc. Here the domains have the shapes that are 
not related to
the Laplacian eigenfunctions. We shall call this transition the 
configurational transition. In Section VII we 
show that the ratio of rectangle sides
$a/b$ at the transition can be 
obtained from the simple condition involving the third Laplacian eigenvalue.
In a natural way our model provides the probabilistic interpretation
of the Laplacian eigenfunctions. 
We would like to emphasize that 
our stochastic model leads to a deterministic limiting distribution,
whereas for example in super-Brownian motion
and Fleming-Viot processes the
limiting distribution has the fractal nature.

The paper is organized as follows. In Section II we 
briefly discuss the discrete analogues of super-Brownian
motion and Fleming-Viot processes. In Sections III and IV
we describe in detail our model. 
The connection between the stationary state of the model
and the Laplacian eigenfunctions
is given in Section V  and  computer simulations
are described in Section VI.
The analysis of the configurational transition and the concluding
remarks are contained in Section VII.

\centerline{\bf II Super-Brownian and Fleming-Viot processes:} 
\centerline{\bf particle
systems with death rate independent of position}

Super-Brownian motion and Fleming-Viot processes
are usually discussed in the continuous time
and space state setting. We will present their discrete
analogues for the purpose of comparison with our own
model introduced in Sections III and IV below.
In the first model we consider particles on the two dimensional
square
lattice. At every time step $t=1,2,3\cdots$, each particle 
either dies or branches off into two offsprings with probability $1/2$.
If the particle branches, both offsprings  occupy the same lattice site as 
the parent particle and then each goes to one of the four
neighbour lattice sites with probability $1/4$. The events are independent
for all the particles in the population. 
Suppose that at time $t=1$ the particle system 
consists of $j$ particles and every particle
is located at $(0,0)$. 
Let $X^j_s$ be a measure-valued process whose
value at time $s$ is defined as follows. The measure
$X^j_s(A)$ of an open subset $A$ of $\R^2$ is equal to the
number of particles at time $t=[s]$
which lie in $\sqrt{j}A$
($[s]$ is the integer part of $s$).
Consider the sequence of processes
$\{X^j_{ju}/j , u\geq 0\}_{j\geq 1}$ where
$s= ju$ and $u$ plays the role of the rescaled time.
This sequence of processes
converges as $j\to\infty$
to a measure-valued diffusion called super-Brownian motion
with the initial state $\delta_{(0,0)}$ (mass
1 concentrated at $(0,0)$).
The limiting distribution of the process has the fractal nature
in dimensions $d \geq 2$. 
At any fixed time, the state of super-Brownian motion
is a measure whose support has Hausdorff dimension 2 [10],
for $d\geq 2$. In other words, the volume occupied by the
particles with the linear size of the system, $L$, 
scales as $L^2$ irrespective of $d\geq 2$ ($L \to 0$).
For $d =1$, the limiting distribution of the process at a
fixed time has a continuous density.

The second model differs from the first one in that
the population size is fixed
and equal to $j$. The dynamics are now the following.
First suppose that $k = 1,2,\dots, j-1$ and $n\geq 1$.
In order to obtain the state of the process at time $t=nj+ k+1$
from that at $t=nj+k$, we choose randomly
one particle and kill it. Next, another particle is
chosen from the surviving ones and it
branches into two offspring which occupy the same
lattice site as the parent particle.
If $t=nj$ then we obtain the new configuration at time
$t= nj+1$ by letting
each of the particles move to one of the $4$
nearest sites on the lattice, with probability $1/4$,
independent of all other particles.
We renormalize the system in order to obtain a continuous limit.
Suppose that at time $t=1$ the system 
consists of $j$ particles located at $(0,0)$. 
Let $X^j_s$ be a measure defined as in the first model, i.e.,
the measure
$X^j_s(A)$ of an open subset $A$ of ${\bf R}^2$ is equal to the
number of particles at time $t=[s]$ which lie in $\sqrt{j}A$.
Then the sequence of processes
$\{X^j_{j^2u}/j , u\geq 0\}_{j\geq 1}$ ($s=j^2u$)
converges as $j\to\infty$
to a measure-valued diffusion called the Fleming-Viot process [6, 11]
with the initial state $\delta_{(0,0)}$.
This process has the same fractal nature as the super-Brownian motion.
More precisely, the state of the Fleming-Viot process
at a fixed time is a measure which has the same Hausdorff
dimension (and other fractal properties like Hausdorff
measure, packing measure, etc.) as super-Brownian motion.
The Fleming-Viot process is the super-Brownian motion
when the latter is conditioned to have a constant
total number of particles.

Recently there has been growing interest in models 
incorporating dependence of the motion of individual
particles on the current
configuration [11,12].
We propose to study a model with a constant population
size in which particles can die and branch. In our
model, the death of a particle will depend on its location and thus it 
differs from the two aforementioned models.
The simplest case is when a particle dies if and
only if it moves to a set of the designated sites on the lattice.

\centerline{\bf III Particle system with death depending on position}

We fix a connected subset $D_\eps$ of 
the square lattice with the mesh size $\eps$, denoted
$(\eps\Z)^2$. The particles
in our model die if they move outside $D_\eps$, so
$D_\eps$ plays the role of the state space. 
The number of particles is fixed and equal to $j$.
Transitions from the state of the system
at time $t=k$ to that at time $t=k+1$ may be described
as follows. First
each of the particles goes to one of the $4$
nearest sites on the lattice $(\eps\Z)^2$, with probability $1/4$,
independent of all other particles.
Then all particles which are outside $D_\eps$ die.
An equal number of particles is chosen uniformly
from among the surviving particles. Each of the chosen
particles splits into two offspring which occupy the same
site as the parent particle. Hence, the number of particles
in our model is constant between generations.

Fix some open connected set $D\subset \R^2$
and let $D_\eps = D \cap (\eps\Z)^2$.
Suppose that at time $t=1$ each of the $j$ particles 
occupy sites in $D_\eps$.
Let  $X^{j,\eps}_s$ be the measure valued process whose
value at time $s$ is defined as follows. The measure
$X^{j,\eps}_s(A)$ of an open subset $A$ of $\R^2$ is equal to the
number of particles which are in $A$ at time $[s]$.

The qualitative long time behavior of our system is much different from
that in the case of the super-Brownian motion
or Fleming-Viot process. 
A typical particle configuration in both models
discussed in the previous section has a fractal nature.
Rigorously speaking, the limiting continuous models
are measure-valued diffusions whose states are measures
supported on fractal sets [11].
In our model, increasing the number of particles $j$
and decreasing the mesh $\eps$ of the lattice results, in the long run,
in a stable distribution
which is a suitably normalized first eigenfunction
of the Laplacian on $D$ with zero boundary values.
In other words, if $f(x,y)$ denotes the first eigenfunction
of the Laplacian with zero boundary values in $D$
then $\lim X^{j,\eps}_{s/\eps^2} (dx,dy)/j = cf(x,y)dxdy$
or, more precisely,
$\lim X^{j,\eps}_{s/\eps^2} (A)/j = \int_A cf(x,y)dxdy$
for every open set $A\subset \R^2$, 
where $0<c<\infty$ and
the limit is taken as $\eps\to0$, $j\to\infty$
and $s\to\infty$.
Here $c = 1 / \int_D f(x,y) dxdy$.

One physical interpretation of the first eigenfunction
is that it represents the probability distribution, 
after a long time delay, for a randomly moving
particle conditioned to stay within
the domain [13]. The normalization of the eigenfunction
is necessary to make the total probability equal to 1
(in the case of probabilistic interpretation)
or to make its integral equal to the total mass
of particles in our model (we normalize the mass by dividing
the measure $X^{j,\eps}_{s/\eps^2}$ by $j$).

We offer a heuristic argument showing the convergence
of distributions in our model. The remarks are not meant
to be a rigorous proof --- that does not seem to be
trivial and will be the subject of a forthcoming paper [14].
Notice first that because of the diffusive scaling 
$x\rightarrow\eps x$ and $s\rightarrow\eps^{-2}s$, each particle, 
in the limit $\eps\rightarrow 0$, executes a Brownian motion in $D$
with a jump, upon exiting $D$, to a point occupied by a fellow particle 
chosen uniformly at random. Second, since particles interact only
through the boundary of $D$ by a random choice from the remaining 
particles, the equal time
pair correlations are inversely proportional to the total particle number $j$,
and therefore the particles are uncorrelated in the limit 
$j\rightarrow\infty$. Thus the 
limiting measure $X_s=\lim_{\eps, j} X^{\eps, j}_{s/\eps^2}/j$
exists and is deterministic. Let us express this limit via its density
$X_s(A)=\int_A\xi(s;x,y)dxdy.$ Since all particles reside in $D$ it follows 
that $\xi(s;x,y)\geq 0$ and it vanishes for points on the boundary of $D$. 
Now the average
exit time of a typical particle from $D$ is the reciprocal of 
$\lambda_1$,
the first eigenvalue of the Laplacian in $D$ with zero boundary 
values [13],
which shows that the per particle rate at which jumps take place is exactly
$\lambda_1$. Thus the density $\xi(s;x,y)$ is the solution of a heat flow 
problem in $D$ with a heat source of strength 
$\lambda_1 \xi(s;x,y) $
and absorbtion at the boundary, i.e.,
$$\prt\xi /\prt s + \triangle\xi=\lambda_1\xi.\eqno(3.1)$$ 
Here,
$\triangle =-1/2(\prt^2/\prt x^2 + \prt^2/\prt y^2)$ is the Laplacian.
As $s\rightarrow\infty$, the density converges to a solution of the stationary 
problem for which the normalized first eigenfunction is the unique 
non-negative solution having total integral 1.

\centerline{\bf IV Multi-type particle system}

The first eigenfunction of the Laplacian with zero
boundary values already has a natural probabilistic 
interpretation [13]
and the model described in the previous section
provides a new one.
It seems that so far the higher eigenfunctions
do not have a natural probabilistic interpretation.
A model described in this section may be a first step
towards such an interpretation.

Fix a connected subset $D_\eps$ of 
the square lattice $(\eps\Z)^2$ with the mesh size $\eps$.
In this model, each particle will reside in $D_{\eps}$ and
have one of $m$ possible types
${\cal L}_k$, $k=1,2,\dots,m$.
Typically, at each time $t=l$ some particles will be chosen to
split into two offspring. In such a case we
will say that a new offspring was born at time
$t=l$ and if this particle is killed at some later time $t = n$
then we will say that the lifetime of this particle
was ${T} = n-l$.
The transition mechanism of the system, which depends on the positions,
types, and lifetimes of the particles, is the following one. First
each of the particles goes to one of the $4$
nearest sites on the lattice $(\eps\Z)^2$, with probability $1/4$,
independent of all other particles.
Then all particles which moved outside $D_\eps$ are killed.
If a site in $D_\eps$ is occupied by particles of several types, then
two particles of different types are chosen randomly and also are killed.
We repeat the procedure, killing pairs of
particles of different type occupying the same site until
there are no sites in $D_\eps$ with more than one type of particle.
Killed particles will be replaced with new offspring as follows.
For every $k$, 
we choose $n_k$ (to be defined below) 
particles of type ${\cal L}_k$
randomly from among the surviving ones and each of these particles
splits into two offspring of the same type
which then  occupy the same site as the parent particle.
Now we define $n_k$. Let $n_k^1$ be the number
of particles of type ${\cal L}_k$ which died because
they moved outside $D_\eps$. Let $n_k^2$ be the number
of the {\it pairs} of particles which were killed inside $D_\eps$
such that the types and lifetimes of the particles involved
were $({\cal L}_i,{T}_1)$ and $({\cal L}_k,{T}_2)$
and ${T}_1 > {T}_2$ (i.e., the particle with
type ${\cal L}_k$ had a shorter lifetime). 
Let $n_k^3$ be defined just as $n_k^2$ except that
we replace the condition ${T}_1 > {T}_2$
with the condition ${T}_1 =  {T}_2$. Then we set
$n_k = n_k^1 + 2 n_k^2 + n_k^3$.
Note that the total number of particles
in our model is constant between generations but the
number of particles of type ${\cal L}_k$ can vary, for each $k$.

Again, we consider the high density limit distribution
for the system.
Fix some open connected set $D\subset \R^2$,
let $D_\eps = D \cap (\eps\Z)^2$ and assume
that at time $t=1$ all particles 
occupy sites in $D_\eps$. Recall that  we have 
the total of $j$ particles which belong to $m$ different types
${\cal L}_k$.
Let the measure
$X^{k,j,\eps}_s$ of an open subset $A$ of $\R^2$ be equal to the
number of particles of type ${\cal L}_k$
which are in $A$ at time $[s]$.

Fix $m\geq 2$ and $D\subset \R^2$ and let $j\to\infty$,
$\eps\to0$ and $s\to\infty$. In the limit, for every $k$,
the measure $X^{k,j,\eps}_{s/\eps^2}(dx,dy)/j$ will converge to
$c_k f_k(x,y)dxdy$ 
(in other words,
$X^{k,j,\eps}_{s/\eps^2}(A)/j \to \int_A c_k f_k(x,y)dxdy$
for every open set $A\subset \R^2$)
where $0< c_k< \infty$
and $f_k$ is the first eigenfunction of the Laplacian
with zero boundary values on a subdomain $D_k$ of $D$.
Because of the dynamics, particles of different types become
segregated so the subdomains $D_k$ are disjoint and their
union is $D$.

Our transformation rules have been chosen so that
the average lifetimes of particles of different types
are equal in the limit. For if at a certain time
the average lifetime of particles of type ${\cal L}_k$
is smaller than that for type ${\cal L}_n$, the collisions
of the particles of these two types will result in an increase
of the number of particles of type ${\cal L}_k$. This will
imply the growth of the subregion $D_k$ occupied by
particles of type ${\cal L}_k$ and hence their average
lifetime will increase. The opposite will be true for
the particles of type ${\cal L}_n$ and so in the limit
the average lifetimes of all types of particles
will be the same.

The average lifetime of a particle of type ${\cal L}_k$
is equal to the inverse of the first eigenvalue
in $D_k$. Hence, the first eigenvalue for the Laplacian
with zero boundary conditions in $D_k$ is the same
for every $k$, in the limit.

Let $(x,y)$ be a point on the boundary between
between two subregions $D_k$ and $D_n$ and let $N$
be the normal unit vector to the boundary at $(x,y)$
pointing inside $D_k$.
Note that the normal unit vector $\hat N$ at $(x,y)$
pointing inside $D_n$ is the same as $-N$.
Then we must have $\prt c_k f_k/ \prt N = -\prt c_n f_n/\prt(\hat N)$
because the particles of both types are killed on the
boundary at the same rate.

\centerline{\bf V Limit distribution and Laplacian eigenfunctions}

Let $F(x,y)dxdy= F_m(x,y)dxdy$ be equal to the limit
of $X^{k,j,\eps}_{s/\eps^2}(dx,dy)/j$ on $D_k$. In other words,
$F(x,y) = c_k f_k(x,y)$ on $D_k$ and the constants
$c_k$ are such that $\prt c_k f_k/ \prt N = -\prt c_n f_n/\prt(\hat N)$
on the boundary between $D_k$ and $D_n$, where
$N$ is the inward normal vector on the boundary of $D_k$
and $\hat N = -N$.
Hence, $\prt F/ \prt N = -\prt F/\prt(\hat N)$
on the boundary between $D_k$ and $D_n$.

Suppose that $g$ is an eigenfunction for the Laplacian
in $D$ with zero boundary values. The lines where $g$ is
equal to zero are called the ``nodal lines'' and they divide
$D$ into a number of subregions $\wt D_k$. The function
$g$ is differentiable, so we must have
$\prt |g|/ \prt N = -\prt |g|/\prt(\hat N)$
on the boundary between $\wt D_k$ and $\wt D_n$.
Moreover, $|g|$ is the first eigenfunction for the Laplacian
on every subregion $\wt D_k$.
This suggests that $F_m$ may be equal to $|g|$ for some
eigenfunction $g$ of the Laplacian in $D$.

A simple example shows that for some $D$ and $m$,
the limit distribution $F_m$
cannot be equal to $|g|$ for any eigenfunction $g$ in $D$.
This is the case when an odd number of ``nodal lines'' 
for $F_m$ meet at a single
point. The number of nodal lines meeting at one point must
be even for an eigenfunction since the sign of the
eigenfunction in adjacent regions defined by its nodal lines
must alternate. There would be no consistent way
of assigning signs to adjacent regions if an odd
number of them met at an intersection point of nodal lines.
Fig. 1b illustrates
a limit distribution for a system with 3 particle types.
In this case, there are three nodal lines for $F_m$
which meet at one point and consequently $F_m$ cannot be equal to
$|g|$ in this case.

One may ask, then, when the limit distribution $F_m$
for a multi-type particle system corresponds to a higher
eigenfunction. We concentrated our efforts on one particular
class of domains, namely rectangles $D$ because
in this case, the eigenvalues
and the corresponding eigenfunctions can be explicitly calculated.

Let $D = \{(x,y) \in \R^2: 0 < x < a, 0<y<b\}$.
Then all eigenvalues of the Laplacian in $D$ with
zero boundary values are given by
$\lambda_{j,k} =\pi^2 [(j/a)^2 + (k/b)^2]$, where $j$ and $k$ are
arbitrary integers greater than 0 [15]. The eigenfunction
corresponding to $\lambda_{j,k}$ has the form
$f_{j,k}(x,y) = \sin( (j\pi/a) x) \sin((k\pi/b)y)$.
It may happen that $\lambda_{j_1,k_1} = \lambda_{j_2,k_2}$
even though $j_1\ne j_2$ and $k_1 \ne k_2$ but this is
possible for only a countable number of side ratios
$r = b/a$. We can also write $\lambda_{j,k}$ as
$(\pi/a)^2 (j^2 +(k/r)^2)$. 

It is intuitively clear that when the number of types
of particles $m$ is constant but the side ratio of $D$ 
is very large then
particles of different types will
occupy $m$ rectangles arranged in a linear order (see for example Fig.1a).
We will call this arrangement ``elementary.'' It corresponds
to an eigenfunction $f_{j,k}$ of the Laplacian with 
either $j=1$ or $k=1$.
This effect is due to the tendency of different populations
to separate and an elementary configuration seems to be
a natural way to achieve maximum separation.
It is not so clear what happens when the side ratio
is moderate.
When $m$ is fixed, say $m=3$,
and the side ratio is close to 1, we obtain in computer
simulations a configuration 
illustrated in Fig.1b which does not correspond
to any eigenfunction. We determined by simulation the critical
side ratio at which we
observe the transition between the elementary configuration
and a configuration which does not correspond
to any eigenfunction.

\centerline{\bf VI Computer simulations}

Further discussion of the limiting distributions
and eigenvalues will be illustrated by computer simulations
so we make a digression to explain our figures.
In all simulations we took $D$ to be a rectangle.
The figures show the regions $D$ and the boundaries
between the subregions occupied by different
particle types. All simulations were done
for rectangles $D$ with sides $b=100$ and $100<a<300$. 
Because of memory constraints,
the results of the simulations were compressed in the following way.
Every region $D$ was divided into a number of small identical
rectangles, usually with side lengths between 5 and 10.
The numbers of particles of different type were found
in every small rectangle and the rectangle was declared of type
${\cal L}_k$ if the number of particles of this type was
the greatest of all particle types. Only rectangles close to 
the boundaries between $D_k$'s contained different particles types.
In our simulations,
almost all other rectangles contained only one type of particles.

We have simulated long time behavior of the system
in rectangles of different side ratios with 100,000 particles.
Most simulations ran for 150,000
or 200,000 timesteps.
The starting configurations included ``elementary
configurations,'' other configurations with polygonal
separating lines and totally random configurations.
We used various initial proportions of different particle
types.
We did simulations with $m=3,4$ and $5$ particle types.
In each case we determined the critical side ratio
$r_m=a/b$ at which we observed a transformation
of the stationary configuration
from the elementary configuration to a configuration
which did not correspond to an eigenfunction.
The simulations were peformed in 20 different rectangles.
Due to time consuming nature of the simulations,
the number of independent samples varied from 1 to 5 per
rectangle.
The final configurations for the segregation phases
were unique and
did not depend on the initial configuration
except when the side ratios were close
to the critical values discussed below. 

When the number of particle types is $m=3$,
the critical side ratio is $1.64\pm 0.01$ (Fig. 1).
The simulations starting from various initial distributions
show that the limit distribution is elementary
for the ratio 1.65 and it is not for the ratio 1.62.
In the case of side length ratios 1.63 and 1.64,
the particle configuration had a tendency to preserve
its initial shape if the initial shape was as in Fig. 1a-b.

The results of the simulations are most clear
in the case of 4 particle types. Each of the simulations
was started from an asymmetric configuration.
The critical ratio is $2.26\pm 0.01$. The particle
distributions are given in Fig. 2.

Simulations with 5 particle types (Fig.3) were also started
from asymmetric distributions. In this case,
the critical side ratio is  $2.85\pm 0.01$.

An ``asymmetric'' initial configuration is illustrated
in Fig. 4.

\centerline{\bf VII Configurational transition and Laplacian eigenvalues}

We will argue that
the critical side ratios obtained from the computer simulations 
match exceptionally well the critical rectangle side ratios
for the following problem.
{\sl When is it true that the elementary configuration
with $m$ subregions corresponds to the $m$-th
eigenfunction? }
We order the eigenfunctions according to the their eigenvalues,
i.e., the $m$-th eigenfunction corresponds to $m$-th
smallest eigenvalue.

Recall the formulae for the eigenvalues of the Laplacian
given in Sect. 3. We have 
$\lambda_{j,k} = (\pi/a)^2 (j^2 +(k/r)^2)$ for a rectangle
with sides equal to $a$ and $b$ and side ratio $r=b/a$. 
The elementary configuration is defined by the eigenfuction
corresponding to $\lambda_{1,m}$. Whether $\lambda_{1,m}$
is the $m$-th eigenvalue depends only on $r$ (it does
not otherwise depend on the values of $a$ and $b$).
Note that $\lambda_{1,k} < \lambda_{1,m}$ for $k<m$
so $\lambda_{1,m}$ is the $m$-th eigenvalue if and only if
$$\lambda_{1,m} < \lambda_{2,1}.\eqno(7.1)$$ 
This is
equivalent to (section V)
$$1^2 +(m/r)^2 < 2^2 +(1/r)^2.\eqno(7.2)$$
We take $m=3,4,5$ and solve this equation for $r$
to obtain the following critical side ratios $r_m$:
$r_3=\sqrt{8/3} \approx 1.63$,  
$ r_4 = \sqrt{5} \approx 2.24$,
$ r_5  = 2^{3/2} \approx 2.83$.

Since our simulations were done on a discrete lattice,
the critical side ratio values calculated for the 
rectangle $D$ in $\R^2$ are only approximate.
Eigenfunctions for the discrete Laplacian
on a rectangle 
$D = \{(x,y) \in \Z^2: 1 \leq x \leq a, 1 \leq y \leq b\}$ 
are given by $\wt f(x,y)=g(x)h(y)$ where $g$ and $h$ satisfy
$g(0) = g(a+1) = 0$, $h(0) = h(b+1) = 0$, and
$$\eqalignno{
g(x-1) - 2 g(x) + g(x+1) &= -\wt\lambda^x g(x), \quad 1 \leq x \leq a,\cr
h(y-1 ) - 2 h(y) + h(y+1) &= -\wt\lambda^y h(y), \quad 1 \leq y \leq b.
}$$
Then $\wt\lambda = \wt\lambda^x + \wt\lambda^y$ is the eigenvalue
corresponding to the eigenfunction $\wt f(x,y)=g(x)h(y)$. 
If $g$ changes sign $j-1$ times and $h$ changes the sign $k-1$
times then $\wt\lambda = \wt\lambda_{j,k}$ is a discrete
analog of $\lambda_{j,k}$.
We have the following explicit formulae for the eigenfunctions 
and the corresponding eigenvalues.
$$\eqalign{
g_j(x) &= \sin(j\pi  x/(a+1)), \cr
h_k(y) &= \sin(k\pi  y/(b+1)), \cr
\wt\lambda_j^x &= 2 (1-\cos (j\pi  / (a+1))), \cr
\wt\lambda_k^y &= 2 (1-\cos (k\pi  / (b+1))). }
$$
The discrete analogues of inequalities (7.1) and (7.2) are
$$\wt\lambda_{m,1} < \wt\lambda_{1,2}$$ 
and
$$\cos (m\pi  / (a+1)) + \cos (\pi  / (b+1))
> \cos (\pi  / (a+1)) + \cos (2\pi  / (b+1)).$$
In the case $b=100$, the critical values for $a$ in the
last inequality are in the following intervals,
$$\eqalign{
163 < a & < 164, \qquad m = 3, \cr
224 < a & < 225, \qquad m = 4,  \cr
284 < a & < 285, \qquad m = 5.}
$$
These values
match very well the critical side lengths discussed in the 
previous section.

It is 
quite intriguing that the configuratinal transition takes place for
side ratio related to the eigenvalue of the Laplacian (Eq(7.1-2)).
It would be interesting to find an explanation for this phenomenon.
We hope that our results will be useful in the future studies of the
population dynamics.

\centerline{\bf Acknowledgements}

This work was supported in part by the NSF grant DMS 9322689 and
KBN and FWPN grants. 

\centerline{\bf References}
\bigskip

\item{[1]} Berlekamp, E.R., Conway, J.H., and Guy, R.K. 1982
{\it Winning Ways for your Mathematical Plays} (Academic Press, New
York vol2)

\item{[2]} Wolfram, S. 1986 {\it Theory and Application of Cellular
Automata}, World Scientific, Singapore

\item{[3]} Sales T.R.M. 1993 Phys.Rev.E {\bf 48} 2418

\item{[4]} Lewenstein M., Nowak, A., and Latan\'e, B. 1992 Phys.Rev. A {\bf 45}
763

\item{[5]} Strobeck C. and Morgan K. 1978 Genetics {\bf 88} 829 

\item{[6]} Ethier S.N. and Kurtz T.G. 1993
SIAM J. Control Optim., {\bf 31} 345

\item{[7]} Zhou J., Murthy G. and S. Redner 1992 J.Phys. A {\bf 25} 5889

\item{[8]} Burlatsky S.F. and Pronin K.A. 1989 J.Phys. A {\bf 22}, 531

\item{[9]} Murray J.D. 1989 {\it Mathematical Biology},  (Berlin-Springer)

\item{[10]} Mandelbrot B.B. 1982 {\sl The Fractal Geometry of Nature}  
(Freeman \& Co., New York)

\item{[11]} Dawson D.A. 1993 {\sl Measure-valued Markov processes} in
``Ecole d'ete de probabilites de Saint-Flour XXI, 1991,''
Lecture Notes in Mathematics, v. 1541, P.L. Hennequin (ed.),
pp. 2--260 ( Springer, New York)

\item{[12]} Perkins E. 1992 Prob. Th. Rel. Fields,
{\bf 94} 189

\item{[13]} Port S. and  Stone C. 1978 {\sl Brownian Motion
and Classical Potential Theory} (Academic Press, New York)

\item{[14]} Burdzy K. and March P., forthcoming paper

\item{[15]} Courant R. and  Hilbert D. 1953 {\sl Methods of Mathematical
Physics} vol. 1 Sec. V.5.4, (Interscience, New York)

\centerline{\bf Figure captions}

{\bf Figure 1}. Nodal lines for stationary distribution of 
particles  
with $3$ particle types. 
Each region, separated by solid lines is occupied by only one type
of particles. ({\bf a}) The side ratio $r_3=a/b=1.64$; 
elementary configuration corresponding to the third
Laplacian eigenfunction
({\bf b}) $r_3=1.63$;
configuration close to the transition point (non-elementary configuration)
({\bf c}) $r_3=1$;
configuration far from the transition point (non-elementary
configuration).

{\bf Figure 2}.  Nodal lines for stationary distribution of 
particles
with $4$ particle types. 
Each region, separated by solid lines is occupied by only one type
of particles. ({\bf a}) The side ratio $r_4=a/b=2.27$; 
elementary configuration corresponding to the fourth
Laplacian eigenfunction
({\bf b}) $r_4=2.24$;
configuration close to the transition point (non-elementary configuration)
({\bf c}) $r_4=1$;
configuration far from the transition point (non-elementary
configuration).

{\bf Figure 3}. Nodal lines for stationary distribution of 
particles
with $5$ particle types. 
Each region, separated by solid lines is occupied by only one type
of particles. ({\bf a}) The side ratio $r_5=a/b=2.88$; 
elementary configuration corresponding to the fifth
Laplacian eigenfunction
({\bf b}) $r_5=2.84$;
configuration close to the transition point (non-elementary configuration)
({\bf c}) $r_5=1$;
configuration far from the transition point (non-elementary
configuration).

{\bf Figure 4}. An ``asymmetric'' initial configuration
with 4 particle types. Configurations of this type were
used as initial configurations for many simulations.

\end